\documentclass[journal]{IEEEtran}
\usepackage{subfiles}
\pdfoutput=1

\ifCLASSINFOpdf
  \usepackage{graphicx}
  \graphicspath{ {Graphics/}{../Graphics/} }
  \DeclareGraphicsExtensions{.pdf,.jpg,.jpeg,.png,.eps}
\else
\fi
\usepackage{amsmath}
\usepackage{bbm}
\usepackage{xfrac}
\usepackage{bigints}
\usepackage{amssymb}
\usepackage{mathtools}
\usepackage{cancel}
\usepackage{empheq}

\newcommand{\E}{\mathbb{E}}
\newcommand{\R}{\mathbb{R}}
\newcommand{\Lagr}{\mathcal{L}}

\usepackage{xcolor}
\usepackage{pifont}
\definecolor{darkgreen}{rgb}{0, 0.9, 0}
\definecolor{myred}{rgb}{0.9, 0, 0}
\newcommand{\greencheck}{{\color{darkgreen}\checkmark}}
\newcommand{\redxmark}{{\color{myred}\times}}
\DeclareRobustCommand
  \mydots{\mathinner{\cdotp\mkern-3mu\cdotp\mkern-3mu\cdotp}}
\makeatletter
\newcommand*{\inlineequation}[2][]{%
  \begingroup
    \refstepcounter{equation}%
    \ifx\\#1\\%
    \else
      \label{#1}%
    \fi
    \relpenalty=10000 %
    \binoppenalty=10000 %
    \ensuremath{%
      #2%
    }%
    ~\@eqnnum
  \endgroup
}
\makeatother

\usepackage{makecell}
\usepackage[math]{cellspace}
\usepackage{array}
\usepackage{stfloats}
\begin{document} %%%%%%%%%%%%%%%%%%%%%%%%%%%%%%%%%%%%%%%%%%%%%%%%%%%%%%%%%%%

\title{Contingency Model Predictive Control\\ for Linear Time-Varying Systems}

% a structure at a ~ this keeps a name from being broken across two lines.
\author{John~P.~Alsterda and J.~Christian~Gerdes% <-this % stops a space
\thanks{J.P. Alsterda is with the Department of Mechanical Engineering, Stanford University, Stanford, CA, 94305 USA (email: \mbox{alsterda@stanford.edu}).}% <-this % stops a space
\thanks{J.C. Gerdes is with the Department of Mechanical Engineering, Stanford University, Stanford, CA, 94305 USA (email: \mbox{gerdes@stanford.edu}).}%
\thanks{Manuscript received 23 February 2021; revised DD Month YYYY. This work was supported in part by Ford Motor Company, Dearborn, MI, 48126 USA.}}%

% you are supposed to have spaces between the names.

\markboth{IEEE TRANSACTIONS ON CONTROL SYSTEMS TECHNOLOGY,~Vol.~\#, No.~\#, Month~YYYY}%
{Contingency Model Predictive Control for Linear Time-Varying Systems}
% The only time the second header will appear is for odd numbered pages
% * You probably will NOT include author names 4 peer review papers *
% Can use \ifCLASSOPTIONpeerreview for conditional compilation if desired
% If you want to put a publisher's ID on the page, do it like this:
% Needed in second column of first page if using \IEEEpubid: \IEEEpubidadjcol

% use for special paper notices: \IEEEspecialpapernotice{(Invited Paper)}

\maketitle % make the title area

% No math, special symbols or citations in abstract. 150-250 words.
\begin{abstract} 

We present Contingency Model Predictive Control (CMPC), a motion planning and control framework that optimizes performance objectives while simultaneously maintaining a contingency plan -- an alternate trajectory that avoids a potential hazard. By preserving the existence of a feasible avoidance trajectory, CMPC anticipates emergency and keeps the controlled system in a safe state that is selectively robust to the identified hazard. We accomplish this by adding an additional prediction horizon in parallel to the typical Model Predictive Control (MPC) horizon. This extra horizon is constrained to guarantee safety from the contingent threat and is coupled to the nominal horizon at its first command. Thus, the two horizons negotiate to compute commands that are both optimized for performance and robust to the contingent event. This article presents a linear formulation for CMPC, illustrates its key features on a toy problem, and then demonstrates its efficacy experimentally on a full-size automated road vehicle that encounters a realistic pop-out obstacle. Contingency MPC approaches potential emergencies with safe, intuitive, and interpretable behavior that balances conservatism with incentive for high performance operation.

\end{abstract}

 %%%%%%%%%%%%%%%%%%%%%%%%%%%%%%%%%%%%%%%%%%%%%%%%%%

% Note: keywords not normally used for peer review papers.

% For peer review papers, you can put extra info on cover pg as needed:
% \fi
%
% For peer review papers, this IEEEtran cmd inserts a page break &
\IEEEpeerreviewmaketitle

\graphicspath{ {Graphics/}{../Graphics/} }

\section{Introduction} \label{section:intro}

% If the first word consists of a single letter:
% \IEEEPARstart{A}{demo}
% If you need the single drop letter followed by normal text:

% First start describing the problem:
%   What is a contingent event?
%   Why are they difficult plan for?

%   Does anyone describe the realm of uncertainties for AVs?

\IEEEPARstart{U}{ncertainty} remains a fundamental problem for automated control systems, especially as the technology expands into difficult applications such as vehicle automation -- building the self-driving car. Uncertainty can take many forms, ranging from measurement accuracy to environmental conditions \cite{Ramirez2012ASystems}. Each type presents its own features and challenges, and therefore potentially warrants its own solution. This article proposes a control architecture to tackle a class of uncertainties called contingencies.

 A contingency is a future event or circumstance that is possible but cannot be predicted with certainty. An actuator may fail, a sensor could malfunction, or an obstacle may emerge. Two factors distinguish contingent events from other uncertainties. First, their risk of occurrence is situational and recognizable by context or sensors. Second, their high salience demands anticipation and the discrete focus of a dedicated safety plan. The control design we propose is tailored to these factors.
 
While our formulation is generalized to any linear system, the experiment we perform and examples referenced herein are from the domain of automated vehicles (AVs), an application for which contingencies are frequent and diverse. When released onto public roads, AVs must handle the surprises that human drivers tackle routinely. They should, for example, avoid children who may run into the street and maneuver safely when encountering an icy surface \cite{HighwayTrafficSafetyAdministration2018AScenarios}.

Contingency planning has been used in industry, government, and military applications, involving the preparation and maintenance of an alternative course of action to meet an unexpected situation \cite{Bloom1994ScenarioPlanning}. For control engineering, planning an explicitly alternative trajectory distinguishes this strategy from many robust systems, which typically create one plan to satisfy all possible outcomes. Robust approaches can be impractically conservative when uncertainties are large and numerous, as is the case driving on legal roads \cite{Mayne2014ModelPromise}.

\begin{figure}[!t]
    \centering %\frame{
	\includegraphics[width=\linewidth,trim={10.7cm 9.45cm 10.9cm 9.1cm},clip]{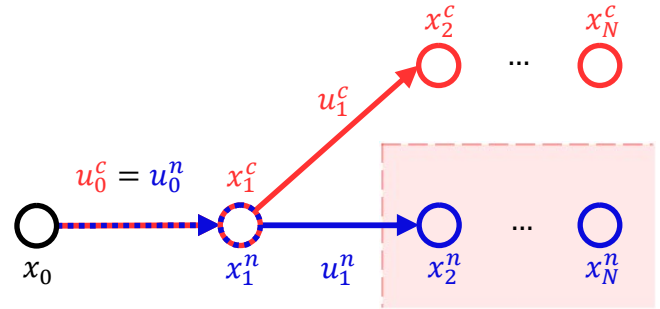}%LBRT
    \caption{Contingency MPC prediction horizons: Nominal horizon (blue) focuses on performance. Contingency horizon (red) must obey the contingent constraints (red-shaded box). The first input, $u_0$, is shared between the horizons, optimized to meet both nominal and contingent objectives.}
    \label{flowchart}
\end{figure}

Contingency planning can be appropriate when a hypothetical situation is markedly different than nominal conditions, entailing new objectives and constraints. For instance, an obstacle in an AV's path presents an obvious change to its constraints: avoid collision. But objectives may also differ; nominal goals such as passenger comfort, efficiency, and time-optimality are not relevant in emergency. Rather, highly dynamic maneuvers may be tolerable and required to maintain safety. 

Another potential contingency is an abrupt change in the road surface. When driving in the snow, for example, a vehicle must be prepared for ice and loss of traction. These two surfaces present different constraints in the form of physics and dynamics equations, but also justify differing objectives. On the snow we seek tight path-tracking performance, but we can ease or disregard this goal in emergency. Contingency planning allows constraints and objectives to be chosen individually for each circumstance.

As our investigation will demonstrate, planning with only a single trajectory can conflate objectives and produce a controller that maintains smooth avoidance maneuvers from unlikely events, at the expense of nominal performance. Contingency planning, in contrast, can facilitate rapid response to sudden events with less modification to nominal operation.

This article is organized as follows: Section~\ref{section:related} places our Contingency Model Predictive Control (CMPC) algorithm in the context of existing techniques to mitigate uncertainty, including other contingency planners and Model Predictive Control (MPC) algorithms. Section~\ref{section:formulation} presents a general CMPC formulation for linear time-varying systems. In Section~\ref{section:toy} we simulate CMPC on a toy problem to explore its properties in comparison to a simple Robust MPC. Section~\ref{section:cmpc4av} develops a CMPC optimization for an automated road vehicle, and Section~\ref{section:results} presents the experimental results of that controller navigating a full-size automated vehicle around a pop-out obstacle, a scenario relevant to leaders in self-driving car development \cite{Tam2020ProactiveDriving}. Finally, Section~\ref{section:discussion} reflects on the current capabilities and limitations of CMPC, and proposes future steps for development.

\section{Related Work} \label{section:related}

An early application of contingency planning in control engineering was high-level routing for robotic vehicles. Linden and Glickaman modified the A* dynamic programming algorithm to choose routes which considered risk of obstruction at bottlenecks \cite{Linden1987ContingencyVehicle}. NASA further developed the strategy for a planetary rover \textit{Contingency Planner/Scheduler}, which planned for the possibility of low battery, instrument failure, or terrain traversal failure \cite{Washington1999AutonomousExploration}. Meuleau and Smith later presented a contingency planner to optimize a rover's daily activities over the belief space, casting the problem as a Partially Observable Markov Decision Process (POMDP) \cite{Meuleau2003OptimalPlanning}.

Hardy \textit{et al.} and Salvado \textit{et al.} revisited contingency planning, narrowing the problem from routing to path-planning \cite{Hardy2013ContingencyVehicles}\cite{Salvado2016ContingencyVehicles}. They proposed and simulated optimizations to navigate probabilistic obstacles, such as an oncoming vehicle that may turn across their intended path. By computing state-trajectories with shared initial segments, their planned movements were prepared for either outcome.

The algorithm we propose advances contingency planning by narrowing the problem further, from path-planning down to closed-loop control by using MPC, an optimal control strategy for systems that require advance planning to achieve high performance and avoid hazards \cite{Richalet1978ModelProcesses}. Whereas a path-planner returns a state trajectory only, MPC computes states and the input commands required to achieve them. MPC has two advantages over path planning: First, it integrates planning and control, eliminating the need for a path-following controller and providing straightforward assurance of dynamic feasibility. Second, it expands the range of contingent events that can be considered to include model-based emergencies (i.e. loss of friction, actuator failure, or other model-mismatch).

We form Contingency MPC by augmenting the typical MPC structure with a second prediction horizon whose task is to return a trajectory that safely navigates an identified contingency. As illustrated in Fig.~\ref{flowchart}, one horizon pursues nominal performance while the other maintains a feasible emergency avoidance.

%The red-shaded box represents a contingent hazard that may become active. The nominal horizon is not concerned with the potential hazard, and pursues normal operating objectives such as path tracking, efficiency, or passenger comfort. The contingency horizon, on the other hand, must stay clear of the hazard and is subject to its own cost function objectives -- objectives more appropriate for emergency operation. $u_0$, the only command scheduled to be deployed at the end of each iteration, CMPC never chooses between plans

Within the field of MPC, there exists a rich and growing collection of techniques to mitigate uncertainty. Most fall under two broad categories: Robust MPC (RMPC) and Stochastic MPC (SMPC). In two reveiws, Mayne and Saltik \textit{et al.} discuss several strategies, challenges in computational and conceptual complexity, and directions for future development  \cite{Mayne2016RobustDirection} \cite{Saltik2018AnAspects}.

In RMPC, uncertain parameters' are confined to a set, and a conservative control trajectory is optimized to satisfy system constraints for all possible combinations of these parameters. Thus RMPC is prepared for the most extreme coincidences of bad luck, however unlikely. Often cast as a min-max problem, RMPC is in general non-convex, leading to difficulty assuring consistent real-time operation \cite{Campo1987ROBUSTCONTROL}. To ensure tractability, Sartipizadeh \textit{et~al.} employed an approximate convex hull, and Hu \textit{et~al.} optimized offline \cite{Sartipizadeh2018AHull}\cite{Hu2019AnMPC}.

Tube MPC is an RMPC that seeks higher performance by computing closed-loop policies rather than open-loop trajectories \cite{Blanchini1990ControlDisturbances}. Tubes are often designed offline, but Lopez \textit{et~al.} recently demonstrated online computation \cite{Lopez2019DynamicSystems}. In the abstract, a perfect policy is a complete contingency plan -- ready with an optimal action for every possibility. Existing Tube MPC algorithms, however, focus more so on mitigating disturbances and modeling error than contingent events.

Stochastic MPC approaches uncertain hazards with less conservatism than RMPC by optimizing over risk directly \cite{Saltik2018AnAspects}. In SMPC, uncertain quantities are modeled as statistical distributions, and inputs are calculated to minimize expected cost. Bujarbaruah \textit{et al.} recently formulated an Adaptive Stochastic MPC to learn model uncertainty online \cite{Bujarbaruah2020AdaptiveUncertainty}.

Scenario SMPC offers a structure that is similar to CMPC, a prediction horizon tree formed by sampling from a distribution at each MPC stage. Krishnamoorthy \textit{et al.} recently improved Scenario SMPC decomposition \cite{Krishnamoorthy2018ImprovingAlgorithm}, and Batkovic \textit{et. al} combined Tube and Scenario MPC to navigate multi-modal obstacles \cite{Batkovic2021AObstacles}. As Bloom and Menefee described, scenario plans are similar to but typically broader than contingency plans, encompassing a larger range of possibilities \cite{Bloom1994ScenarioPlanning}. Indeed, Scenario SMPC can prepare for a wide spectrum of possible futures but is unlikely to sample the specific and more extreme events for which contingency planning is designed.

A core element of any SMPC is a disturbance model, a probability distribution or some function to produce random samples \cite{Farina2016StochasticReview}. Some control applications, however, may not be suitable to likelihood modeling. For example, an accurate distribution may not exist for a road's coefficient of friction, or we may not feel comfortable using probability to predict a child's movement. Contingency MPC offers an avoidance strategy that does not depend on a probabilistic model.

This article further develops the CMPC algorithm we first introduced to navigate uncertain friction conditions, including experimental demonstration on an extreme polished-ice surface \cite{Alsterda2019ContingencyVehicles}. Dallas \textit{et al.} recently expanded that formulation to update friction uncertainty online \cite{Dallas2020ContingentEnvironments}, and Ivanovic \textit{et al.} extended CMPC to navigate roadways with multi-modal predictions of pedestrians and other agents.

%CMPC offers an option for control engineers who seek robustness from foreseeable but uncertain events, achieving responsible and practical conservatism by balancing the pursuit of high performance and safe navigation explicitly in a single real-time optimization.

% More contrast to RMPC conservatism: good strat for a nuclear reactor \cite{Eliasi2012RobustPlant}. Cautious when uncertainty is large. Achieves responsible but practical conservatism.

%Whereas traditional RMPC would aim to maintain comfortable \& efficient emergency maneuvers, CMPC preserves only a feasible avoidance trajectory.

% Monte Carlo MPC? Calculates a single input trajectory, good for all. Like Q-MPD. POMDP in which world becomes observable after next time step. Contingencies drawn from random disturbance distribution.

 %%%%%%%%%%%%%%%%%%%%%%%%%%%%%%%%%%%%%%%%%%%%%%

\section{Contingency MPC Formulation} \label{section:formulation}

Classical MPC is a receding horizon optimal control technique that pursues performance and safety by minimizing a cost function while subject to explicit constraints such as plant dynamics and state or input boundaries \cite{Richalet1978ModelProcesses}. At each time-step, MPC calculates a prediction horizon -- a state trajectory with the open-loop input commands required to achieve it. Upon completion of each iteration, the horizon's first command $u^0$ is deployed, and a new optimization commences. In this article, we consider a fast convex formulation, in which the next optimization will converge before $u^0$ from the previous time-step expires. Therefore, only $u^0$ from each time-step should ever be actuated. Future references to \textit{deterministic} MPC or a \textit{nominal} prediction horizon also refer to this algorithm.

\subsection{General Contingency MPC}

CMPC is a development from classical MPC, in which the nominal horizon is augmented by an additional horizon -- a contingency plan. Both horizons are optimized together, simultaneously. A contingency plan is an avoidance trajectory which mitigates a contingent event, a potential circumstance identified by perception or some other sub-system.

CMPC's paired prediction horizons are illustrated in Fig.~\ref{flowchart}. The nominal and contingency trajectories each stem from $x^0$, the current measured state. $x^n_{*}$ is the nominal state trajectory, analogous to a classical MPC horizon. These are the states we intend to drive the system through by applying $u^n_{*}$, the nominal input trajectory. $x^c_*$ and $u^c_*$ are the contingency plan, which is subject to unique constraints illustrated by the red shaded keep-out area. These constraints encode the hazard posed by a potential emergency; the nominal horizon does not see them.

Critically, the root inputs $u^n_0$ and $u^c_0$ are constrained to be identical; references to $u_0$ simply refer to these values. The equality condition assures that the $u_0$ ready for deployment is both optimized for performance and robust to the contingency. When the contingency plan forecasts danger, the two objectives must negotiate to find a $u_0$ agreeable to both horizons. Nominal path-following and performance will be sacrificed to ensure contingency states remain safe.

%\begin{figure}%[!t]
%    \centering
%	\includegraphics[width=1.3in]{CMPC_flowchart2.png}
%	% [trim={0cm 0cm 0cm 0cm},clip,scale=.7] LBRT
%    \caption{Contingency MPC prediction horizon. A nominal and emergency trajectory are computed, which share a zeroth action.}
%    \label{flowchart}
%\end{figure}

With this design, our controller is never required to choose between the nominal and contingency plans -- the $u_0$ returned in each iteration is always viable for both trajectories. As the system drives forward in time, there are two possibilities: Most often, the contingent event will not occur. As its possibility fades the contingency constraints recede and the system resumes normal operation. Occasionally, the contingent event will occur; having anticipated this possibility, the system is poised to execute an avoidance maneuver.

Next, we define the CMPC objective function $J$ as the sum of nominal and contingency costs at each of $N$ stages. To provide some notational relief, sub- and super-scripts placed outside parentheses or brackets in this article apply to all variables within, such that $(x,u)_k^n = (x_k^n,u_k^n)$.
\vspace{-0.0 cm}

%\begin{align*} \label{eq:J_gen} \tag{1a}
%    J(x_0) \; &= \; \sum_{k=0}^{N} \quad j(x,u)_k \\
%    &= \; \sum_{k=0}^{N} \quad j^n(x,u)_k^n \; + \; %j^c(x,u)_k^c
%\end{align*} \vspace{-0.0 cm}

\begin{align*} \label{eq:J_gen} \tag{1a}
    J(x_0) \;
    = \; \sum_{k=0}^{N} j(x,u)_k \;
    = \; \sum_{k=0}^{N} j^n(x,u)_k^n + j^c(x,u)_k^c
\end{align*} \vspace{-0.5 cm}

Each horizon is subject to its own cost function, $j^n$ or $j^c$, to keep nominal and contingency objectives independent. This is important because normal operation and emergency maneuvering are fundamentally different. A road vehicle during normal operation, for example, has objectives such as passenger comfort and efficiency that are less relevant during emergency. We can relieve the contingency horizon of these inappropriate costs to focus on preserving a safe avoidance maneuver. By allowing the control engineer to prescribe context-appropriate costs to each horizon, the cost functions can be designed to optimize their intended operational domain. 

The constrained optimization problem statement now follows. The root commands for each horizon must be equal. State transitions follow a dynamics model $f$, which may be different in the nominal and contingency plans (and potentially leads to $x_1^n \neq x_1^c$). Further constraints $g$ may encode state or actuator limits and encode contingency hazards.
\vspace{.3 cm}

$\underset{u^n,u^c}{\text{minimize}} \; J(x_0)$, \hspace{.06 cm} subject to:
\vspace{-.3 cm}

\begin{alignat*}{3}
    &\hspace{.35 cm} u^n_0 &&= u^c_0 = u_0 &&
    \label{eq:u_gen} \tag{1b}
    \\[.3 cm]
    &x^n_{k+1} &&= f^n(x,u)^n_k \hspace{.4cm} &&\forall \hspace{.4cm} k
    \label{eq:dyn_gen} \tag{1c} \\
    \text{and} \quad &x^c_{k+1} &&= f^c(x,u)^c_k \hspace{.4cm} &&\forall \hspace{.4cm} k
    \\[.3 cm]
    &\hspace{.2 cm} g^n(&&x,u)^n \leq 0 \hspace{.4cm} &&\forall \hspace{.4cm} g^n
    \label{eq:leq_gen} \tag{1d} \\
    \text{and} \quad &\hspace{.2 cm} g^c(&&x \, ,u)^c \leq 0 \hspace{.4cm} &&\forall \hspace{.4cm} g^c
\end{alignat*}
\vspace{-0.6 cm}
%\vfill
%\null
%\columnbreak

\subsection{Linear Contingency MPC}
\vspace{0 cm}

Generally, CMPC can be adapted onto any MPC. But for the remainder of this article, we focus on a convex form known as a quadratic program (QP), which has a quadratic objective function and linear (affine, precisely) constraints \cite{Boyd2004ConvexOptimization}. Convex programming allows certain CMPC features to be more easily proved and demonstrated, and yields speedy computations that can run online for real-time experiments. A convex CMPC formulation follows:

First we refine the definition of states and inputs as vectors $\textbf{x} \in \R^{2 \cdot n}$ and $\textbf{u} \in \R^{2 \cdot m}$, respectively, for a controlled system with states and inputs of dimension $n$ and $m$.
\vspace{0.0 cm}

\begin{equation} \label{eq:x_u_lin} \tag{2a}
    \textbf{x}_k =
	\begin{bmatrix*}[c] x^n \\ x^c \end{bmatrix*}_k
    \hspace{.4cm} ; \hspace{.4cm}
    \textbf{u}_k =
	\begin{bmatrix*}[c] u^n \\ u^c \end{bmatrix*}_k
\end{equation}
\vspace{-0.1 cm}

Next we develop a convex objective function $J_{cvx}$. We desire to minimize the expected cost $j$ at each stage:
\vspace{-.1 cm}

\begin{align*} \label{J_exp} \tag{2b}
    J_{cvx}(x_0) \, = \; &\sum_{k=0}^{N} \; \E_{C} \Bigg[ \; j(x,u)_k \; \Bigg] \\[4 pt]
    = \; &\sum_{k=0}^{N} \; P^n \cdot j(x,u)_k^n \, + \, P^c \cdot j(x,u)_k^c
\end{align*}
\vspace{0 cm}

The expectation is expanded over the set of outcomes $C$ with their associated likelihoods. In this article we consider only two outcomes: the contingency occurs (with probability $P^c \in [0, 1]$) or it does not (\mbox{$P^n = 1-P^c$}). The expansion is a now convex combination on $j$. It's critical to note that the safety of CMPC does not depend on the accuracy of assigning $P^c$. CMPC remains robust to the contingent event regardless of $P^c$'s value; a feasible avoidance trajectory will always be maintained. Rather, $P^c$ is a knob that tunes CMPC's focus on nominal performance objectives as it approaches a contingency. Using $P^c$ to separate nominal and contingency costs yields an elegant formulation with intuitive behavior, described further in Section~\ref{section:discussion}.

To complete the objective function, we narrow our choice of $j$ to a weighted 2-norm on $x$ and $u$. Other convex terms are allowed, such as the 1-norm we add in Section~\ref{section:cmpc4av}, but are neglected here for compactness. $Q$ and $R$ are positive semi-definite matrices $\in \R^{n \cdot n} \, \text{and} \in  \R^{m \cdot m}$ respectively.
\vspace{-0.2 cm}

\begin{alignat*}{2} \label{J_lin}
    J_{cvx}(x_0) = & \sum_{k=0}^{N} \; \E_{C} \Bigg[ \; x^\top \! Q \, x + u^\top \! R \, u \; \Bigg]\raisebox{-2.5ex}{\small{\textit{k}}}
    \\[.1 cm]
    = & \sum_{k=0}^{N} \; P^n \cdot \Big( x^\top \! Q \, x + u^\top \! R \, u \Big)^n_k \\[0 cm] \tag{2c}
    & \hspace{.2cm} + \; P^c \cdot \Big( x^\top \! Q \, x + u^\top \! R \, u \Big)^c_k
    \\[.1 cm]
    = & \sum_{k=0}^{N} \; \textbf{x}^\top_k \begin{bmatrix}
        P^nQ & 0 \\[0 cm]
        0  & P^cQ
    \end{bmatrix}
    \textbf{x}_k \\
    & \hspace{.2cm} + \; \textbf{u}^\top_k \begin{bmatrix}
        P^nR & 0 \\
        0  & P^cR
    \end{bmatrix}
    \textbf{u}_k \\
\end{alignat*}
\vspace{-.8 cm}

The resulting cost function is quadratic on the stacked state and input vectors from (\ref{eq:x_u_lin}). With a cost function in hand, the constrained convex optimization problem statement follows: \vspace{.5 cm}

$\underset{\textbf{u}}{\text{minimize}} \; J_{cvx}(x_0)$, \hspace{.06 cm} subject to: \vspace{-0.2 cm}

\begin{equation} \label{eq:u_lin} \tag{2d}
    u^n_0 = u^c_0 = u_0
\end{equation} \vspace{-1.0 cm}

\begin{align*} \label{eq:dyn_lin}
    \textbf{x}_{k+1} =
    &\begin{bmatrix*}[c] x^n \\ x^c \end{bmatrix*}_{k+1}
    \hspace{5 cm} \\[.3 cm]  \tag{2e}
    =
    &\begin{bmatrix*}[c] A^n & 0 \; \\ 0 & A^c \; \end{bmatrix*}_{k}
    \!\! \textbf{x}_k +
    \begin{bmatrix*}[c] B^n & 0 \; \\ 0 & B^c \; \end{bmatrix*}_{k}
    \!\! \textbf{u}_k +
    \begin{bmatrix*}[c] C^n \\ C^c \end{bmatrix*}_{k} \\[.4 cm] =
    &\; \textbf{A}_k \; \textbf{x}_k + \textbf{B}_k \; \textbf{u}_k + \textbf{C}_k
\end{align*} \vspace{-.3 cm}

\begin{equation} \label{eq:leq_lin} \tag{2f}
    \begin{bmatrix*}[c] G^n & 0 \; \\ 0 & G^c \; \end{bmatrix*}_k
    \!\! \textbf{x}_k +
    \begin{bmatrix*}[c] H^n & 0 \; \\ 0 & H^c \; \end{bmatrix*}_k
    \!\! \textbf{u}_k
    \; \leq \; b
\end{equation} \vspace{0.0 cm}

As before, $u^n_0$ and $u^c_0$ must be equal. The dynamics model $f$ is now limited to be affine, implemented with block-diagonal dynamics and input matrices. Lastly, the inequality constraints are limited to affine functions defined here by matrices $G$ and $H$ and offset vector $b$.

To highlight the time-varying capacity of this formulation, note that these matrices may change with each time-step $k$. To demonstrate, we successively re-linearize the nonlinear AV dynamics in Section~\ref{section:cmpc4av} as the system moves among operating points.

%The convex program can be solved quickly online for control systems requiring high-frequency updates. Linear CMPC delivers control inputs $u_0$ which maximize cost function objectives to the greatest extent possible, while maintaining an avoidance trajectory from the identified contingent event.

 %%%%%%%%%%%%%%%%%%%%%%%%%%%%%%%%%%%%%%%%%%

\section{Toy-Problem Simulation} \label{section:toy}%%%%%%%%%%%%%%%%%%%%%%%%%

\begin{figure}[t]
    \centering
	\includegraphics[width=\linewidth,trim={4.8cm 6.5cm 6.2cm 6.5cm},clip]{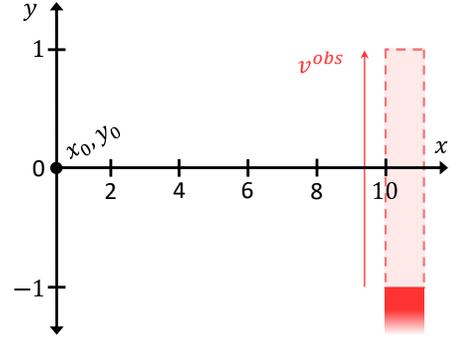}%LBRT
    \caption{Toy problem state-space with the contingent obstacle's path.}
    \label{toy_env}
\end{figure}

To illustrate CMPC's properties and behavior, we pose a toy problem with minimum complexity and simulate the controller's approach. In contrast to a more conservative RMPC, CMPC achieves higher performance as measured by the objective function. We also explore the effect of varying $P^c$, and show how CMPC and RMPC behavior converges as $P^c$ tends toward 100\%.
\vspace{-0.1 cm}

\subsection{System dynamics and control objectives} %%%%%%%%%%%%%%%%%%%%%%%%%

The toy problem state-space is the two dimensional environment shown in Fig.~ \ref{toy_env}, with $x$ and $y$ axes. The controlled system is a point-mass that initially resides at the origin:
\vspace{0.0 cm}

\begin{equation}  \label{eq:toy_x0} \tag{3a}
    x \; , \; y \in \R \quad ; \quad x_0 = y_0 = 0
\end{equation}
\vspace{-0.3 cm}

\noindent When the simulation commences, it moves in the $+x$ direction at a constant speed of $1 \; ^{unit}/_{time-step}$, heading toward the red-shaded contingency region at $x=10$. An obstacle currently lies there at the safe height of $y=-1$, but may spring upward at any time. Fortunately the mass's $y$ position is controllable; its next value can be chosen freely with command input $u$. This movement is captured by the dynamics model:

\begin{equation}  \label{eq:toy_dyn} \tag{3b}
    \begin{bmatrix*}[c] \, x \,\, \\ \, y \,\, \end{bmatrix*}_{k+1} \!\!\!
    = \; \begin{bmatrix*}[c] \, x \,\, \\ \, y \,\, \end{bmatrix*}_k
    + \; \begin{bmatrix*}[c] \, 1 \,\, \\ \, u \,\, \end{bmatrix*}_k
\end{equation} \vspace{-0.1 cm}

\noindent This state-space equation is separable. (\ref{eq:toy_x0}) and (\ref{eq:toy_dyn}) yield the following real-valued equations for the system state at any given time-step $k$: \vspace{-.2 cm}

\begin{equation} \label{eq:toy_dyn2} \tag{3c}
    x_k = k \quad \text{and} \quad
    y_k = \sum_{i=0}^{k-1} u_i
\end{equation} \vspace{-0.1 cm}

The obstacle has the following properties, all known to the controller: It is a hurdle barrier at $x=10$, extending downward to $y=-\infty$. Its height begins at $y^{obs}_{k=0}=-1$, but may spring upward to maximum height of $y^{obs}_{max}=+1$, as indicated in Fig.~\ref{toy_env}. If and when the obstacle triggers, it begins to move upwards at a constant speed of $v^{obs}=0.25 \; ^{units}/_{time-step}$. Looking forward from a current time-step $k$, to when the point-mass arrives at $x=k=10$, the obstacle height will be:
\vspace{-0.1 cm}

\begin{equation} \label{eq:toy_obs_dyn} \tag{3d} \begin{aligned}
    y^{obs}_{k=10}
    &= min( \; y^{obs}_k+ \quad \Delta k \; \cdot \; v^{obs}
    &&, \; y^{obs}_{max} \, ) \\
    &= min( \; y^{obs}_k+(10-k) \cdot 0.25 \hspace{-.2 cm}
    &&, \; 1.0 \; )
\end{aligned} \end{equation} \vspace{-0.4 cm}

The constraint imposed upon the point-mass's $y$ position is then: \vspace{-.3 cm}

\begin{equation} \label{eq:toy_obs} \tag{3e} \begin{aligned}
    y_{k=10} \; &\geq \; y^{obs}_{k=10}
\end{aligned} \end{equation} \vspace{-.2 cm}

The control objective for the point-mass is to safely navigate across the x-axis using the minimum control effort. We encode this with the following cost function: \vspace{-0.1 cm}

\begin{equation} \label{eq:toy_J} \tag{3f}
    J = \min_{\textbf{u}} \sum_{k}^{} \; u_k^2
\end{equation}
\vspace{-0.5 cm}

\subsection{Robust MPC Solution} %%%%%%%%%%%%%%%%%%%%%%%%%%%%%%%%%%%%%%%%%%%%

\begin{figure*}
    \centering
	\includegraphics[width=\linewidth,trim={0.35cm 11.55cm 0.7cm 11.55cm},clip]{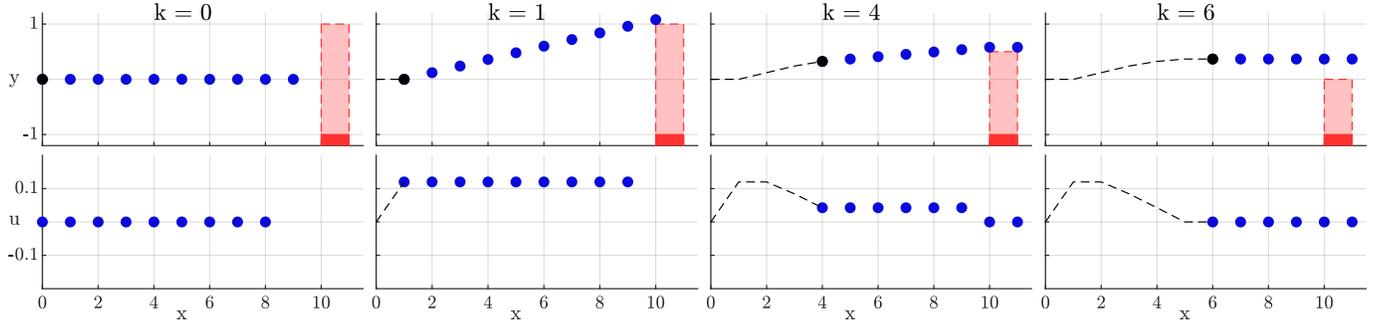}%LBRT
    \caption{RMPC toy problem solution at 4 time-steps. Top row: current state (black), horizon states (blue), obstacle's potential footprint (light red). Bottom row: horizon input commands (blue). This `worst' case algorithm begins deviating immediately, even though the obstacle never pops out.}
    \label{toy_rmpc_soln}
\end{figure*}

\begin{figure*}
    \centering
	\includegraphics[width=\linewidth,trim={0.35cm 11.55cm 0.7cm 11.55cm},clip]{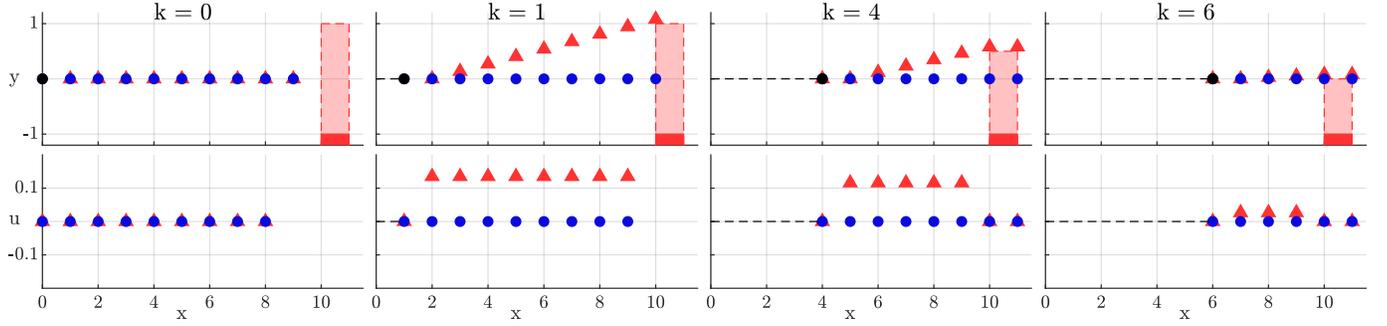}%LBRT
    \caption{CMPC toy problem for $P^c=0\%$. Top row: nominal horizon states (blue) and contingency horizon states (red). Bottom row: nominal horizon inputs (blue) and contingency horizon inputs (red). A contingency plan was maintained, but deviation from $y=0$ was not required because the obstacle did not pop.}
    \label{toy_cmpc_soln}
\end{figure*}

To highlight the behavior of CMPC we compare it to a baseline RMPC, which assumes the worst case evolution of the scene and has a single trajectory like a classical deterministic MPC. Its solution follows. We cast RMPC with a prediction horizon of $N=10$ state tuples $(x,y)$ and inputs $u$. Its cost function takes the form: \vspace{0 cm}

\begin{equation} \label{eq:toy_J_RMPC} \tag{4a}
    J \, = \, \min_{\textbf{u}} \sum_{k=0}^{N}
    \; u_k^2 \, = \, \min_{\textbf{u}} \; \textbf{u}^\top \textbf{u}
\end{equation}
\vspace{-0.0 cm}

It is constrained by the dynamics in (\ref{eq:toy_dyn}) and obstacle in (\ref{eq:toy_obs}), and was solved exactly into the Explicit MPC \cite{Bemporad2002TheSystems} solution presented in the Appendix: % \ref{section:appendix}
\vspace{-0.2 cm}

\begin{flalign} \label{eq:toy_rmpc_sol} \tag{4b}
    u_0 = y^{obs}_{max}/N
\end{flalign}
\vspace{-0.2 cm}

Fig.~\ref{toy_rmpc_soln} shows the RMPC solution at 4 time-steps as it approaches the hurdle, which in this instance never triggers. Assuming the worst case, RMPC begins evasive action immediately when the obstacle comes into view of its horizon at $k=1$. As the mass approaches in time-steps $k=5$ and $k=8$, the red-shaded contingency region recedes, indicating reduction of $y^{obs}_{k=10}$ per equation (\ref{eq:toy_obs_dyn}). In each solution commands are evenly spread out, amortizing their cost over the horizon.

%\begin{figure*}\centering
	%\includegraphics[width=\linewidth]{Graphics/toy_cmpc_plots.eps}\end{figure*}

\subsection{Contingency MPC Solution} %%%%%%%%%%%%%%%%%%%%%%%%%%%%%%%%%%%%%%%
\vspace{0.0cm}

To formulate a CMPC controller, we first concatenate the states and inputs:
\vspace{-0.0 cm}

\begin{equation} \label{eq:toy_y_u} \tag{5a}
    \textbf{y}_k =
	\begin{bmatrix*}[c] y^n \\ y^c \end{bmatrix*}_k
    \hspace{.4cm} ; \hspace{.4cm}
    \textbf{u}_k =
	\begin{bmatrix*}[c] u^n \\ u^c \end{bmatrix*}_k
\end{equation}
\vspace{-0.0 cm}

\noindent The horizontal state $x$ evolves identically regardless of $u$, and is not required to duplicate. The cost function becomes:
\vspace{-0.0 cm}

\begin{equation} \label{eq:toy_J_cmpc} \tag{5b} \begin{aligned}
    J &= \min_{\textbf{u}} \sum_{k=0}^{N} 
    \; P^n \! \cdot \! {u_k^n}^2 + P^c \! \cdot \! {u_k^c}^2 \\[4 pt]
    &= \min_{\textbf{u}} \sum_{k=0}^{N} \hspace{4 pt} \textbf{u}_k^\top \!
    \begin{bmatrix*}[c] \, P^n & 0 \, \\ \, 0 & P^c \, \end{bmatrix*}
    \textbf{u}_k
\end{aligned} \end{equation} \vspace{-0.0 cm}

This quadratic form matches the cost function introduced in equation (\ref{eq:lin_J}), with $Q=0$ and $R=1$. The constrained optimization is then: \vspace{.3 cm}

$\min\limits_{\textbf{u}} \; J(x_0,y_0)$, \hspace{.06 cm} subject to: \vspace{-0.0 cm}

\begin{equation} \label{eq:u0_toy} \tag{5c}
    u^n_0 = u^c_0 = u_0
\end{equation}
\vspace{-0.6 cm}

\begin{equation} \label{eq:dyn_toy_cmpc} \tag{5d}
    \begin{bmatrix*}[c]
        \, x \;\; \\ \, y^n \, \\ \, y^c \,
    \end{bmatrix*}_{k+1} \hspace{-.3 cm}
    = \hspace{.2 cm} \begin{bmatrix*}[c]
        \, x \;\; \\ \, y^n \, \\ \, y^c \,
    \end{bmatrix*}_k
    + \hspace{.2 cm} \begin{bmatrix*}[c]
        \, 1 \;\; \\ \, u^n \, \\ \, u^c \,
    \end{bmatrix*}_k
\end{equation}
\vspace{-0.4 cm}

\begin{equation} \label{eq:toy_obs_cmpc} \tag{5e} \begin{aligned}
    y^c_{k=10} \; &\geq \; y^{obs}_{k=10}
\end{aligned} \end{equation}
\vspace{-0.1 cm}

The CMPC problem also lends to an explicit solution, found in the Appendix with an analytical comparison to RMPC:
\vspace{-.2 cm}

\begin{flalign} \label{eq:toy_cmpc_sol} \tag{5f}
    u_0 = y^{obs}_{max} \frac{P^c}{P^c + N -1}
\end{flalign} \vspace{-0.2 cm}

First, the no-pop scenario is illustrated in Fig.~\ref{toy_cmpc_soln}, with \mbox{$P^c=0\%$} to draw the greatest contrast to RMPC. At $k=1$, the obstacle's path is now visible to the contingency horizon, which charts a path around it. The nominal horizon, however, does not see the potential intrusion and remains at $y=0$. Because $P^c = 0\%$, the contingent commands incur no penalty and CMPC postpones the avoidance to $u_{k>0}$. $J=u_0=0$ for every time-step; the point-mass need not deviate from $y=0$ for an obstacle that does not pop up.

\begin{figure*}
    \centering
	\includegraphics[width=\linewidth,trim={0.35cm 11.55cm 0.7cm 11.55cm},clip]{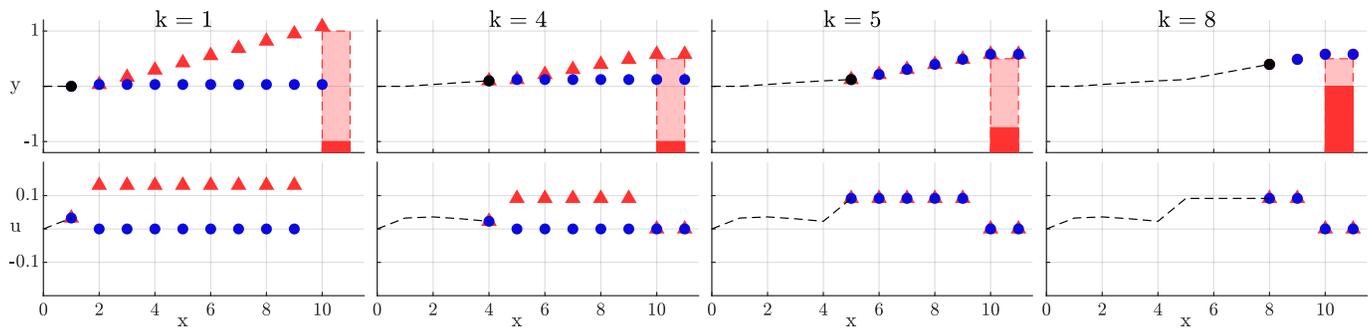}%LBRT
    \caption{CMPC toy problem solution for $P^c=25\%$. The obstacle (red block) begins to pop up at $x=4$. For $k \geq 5$, both horizons see the moving obstacle.}
    \label{toy_cmpc_pop_soln}
\end{figure*}

%\begin{figure*}\centering
	%\includegraphics[width=\linewidth]{Graphics/toy_cmpc_pop_plots.eps}\end{figure*}

\begin{figure}[!b]
    \centering
	\includegraphics[width=\linewidth-1.6cm,trim={6.6cm 7.81cm 6.6cm 7.6cm},clip] {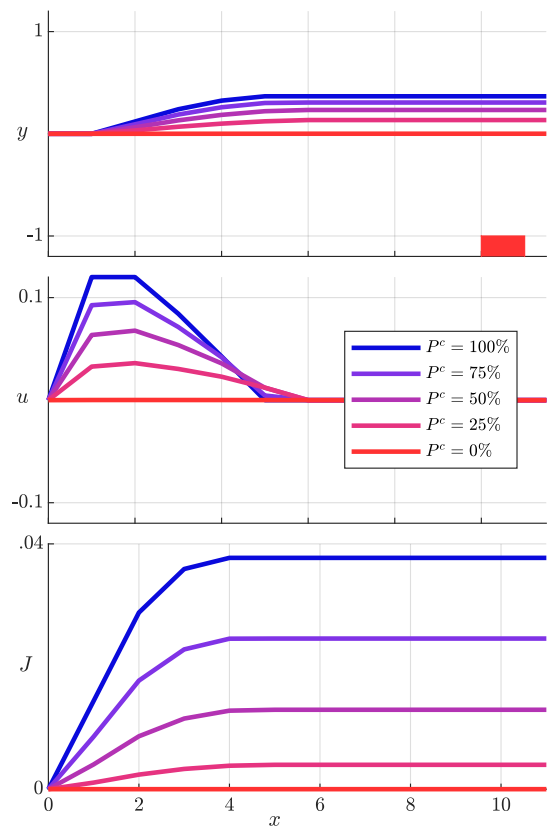}%LBRT
    \caption{Contingency MPC solutions for a range of $P^c$ values.}
    \label{toy_sweep}
\end{figure}

Next, we sweep $P^c$ from 0\% to 100\% for the no-pop scenario. Fig.~\ref{toy_sweep} shows the closed loop CMPC states, inputs, and costs incurred. At 100\%, the behavior is identical to Robust MPC, as expected from equations (\ref{eq:toy_rmpc_sol}) and (\ref{eq:toy_cmpc_sol}). CMPC approaches more aggressively approach as $P^c$ is reduced.
	
%Remember, according to toy dynamics, control input $u_0$ can move $y_1$ upwards to any height; therefore the obstacle remains avoidable for all $k$. In Section~\ref{section:results}, however, CMPC takes protective action even in the absence of positive obstacle detection, to preserve feasibility of avoidance from a more significantly intrusive emergency.

Now the obstacle is allowed to trigger, popping at $k=4$ in Fig.~\ref{toy_cmpc_pop_soln} with $P^c$ set to $25\%$. For $k = 1:4$, CMPC deviates from $y=0$ in anticipation of the potential obstacle. At $k=5$, the obstacle has been moving for one full time-step. The point-mass observes this movement and recognizes the contingency has occurred. The control engineer can design CMPC to respond to activated contingencies in several ways, which we enumerate in Section~\ref{section:discussion}. In this example we chose to alert the nominal horizon to the obstacle's movement; therefore both horizons agree for the remainder the avoidance and the point-mass safely escapes at $x=10$.

\begin{figure}[t]
    \centering
	\includegraphics[width=6.6 cm,trim={6.8cm 10.2cm 6.65cm 10.55cm},clip] {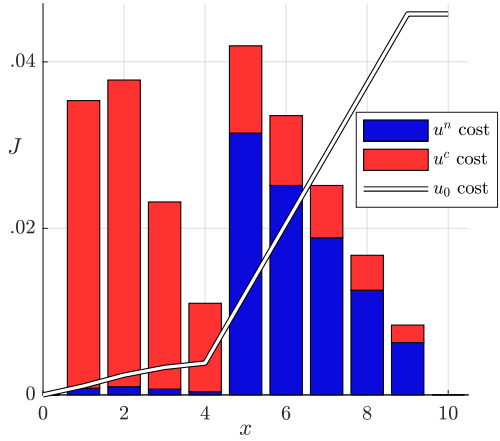}%LBRT
    \caption{Optimal costs for the Fig.~\ref{toy_cmpc_pop_soln} simulation. Blue \& red show CMPC plan costs. The line shows cumulative cost incurred by executing $u_0$ commands.}
    \label{toy_pop_cost}
\end{figure}

%\begin{figure}[b]\centering
	%\includegraphics[width=6.6 cm]{toy_cost_plot.eps}\end{figure}
	
%First, $P^c$ can be rolled towards 100\% as measurement confidence increases. Second, the obstacle's presence can be shared with the nominal horizon, such that both plans now focus on avoidance. Finally, positive recognition could indicate the obstacle is no longer \textit{contingent}, and CMPC's two horizons could be collapsed into to one. Choice among these options is left to the control engineer, whose design will consider the type of circumstance, computational burdens, measurement confidence etc. More discussion of these and other choices in Section~\ref{section:discussion}.

Fig.~\ref{toy_pop_cost} provides a lens into CMPC's objective function. The costs in each iteration are shown as a bar-graph, and the cumulative cost incurred by executing $u_0$ is plotted as a line (penalties on $u^*_{k>0}$ are not actually incurred by the system). For $x=0:4$, avoidance costs are absorbed mostly by the contingency horizon and are minimally present in $u_0$. After $x=4$, the obstacle has been triggered and significant costs are incurred to avoid it. If $P^c$ had been set higher, cumulative $u_0$ cost would be reduced.

\begin{figure}[b]
    \centering
	\includegraphics[width=2.5in]{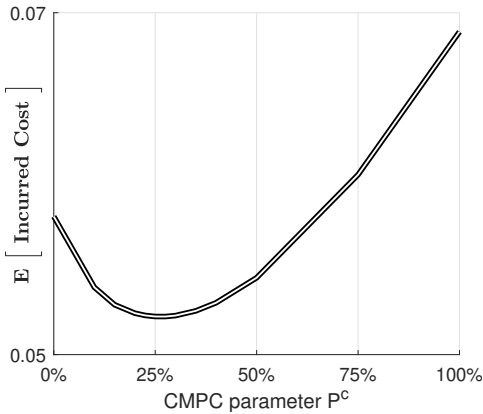}
	%\includegraphics[width=2.5in]{Graphics/toy_expcost-eps-converted-to.pdf}
	% [trim={0cm 0cm 0cm 0cm},clip,scale=.7] LBRT
    \caption{Expected cost incurred as a function of $P^c$ in the toy problem.}
    \label{toy_exp_cost}
\end{figure}

To conclude this toy problem study, we simulated the encounter a large number of times for the entire range of $P^c$ values, triggering the obstacle randomly with probability $10\%$ per time-step. Fig.~\ref{toy_exp_cost} shows the results, indicating that $P^c \approx 25\%$ minimized the expected cost among all possible scenarios (pop-up commences at $k=1,2,...,10$ or not at all). When $P^c$ is set too low, CMPC approaches too aggressively and often requires an abrupt avoidance maneuver. When $P^c$ is too high, CMPC acts too conservatively and avoids the obstacle more than necessary. At $25\%$, CMPC deviates a little bit in each approach, regardless of whether the obstacle triggered. When a pop-out does occur, CMPC is well positioned to escape without incurring too much $u^2$ penalty.

It's interesting to note that the optimal value for $P^c$ is not easily derivable, even for this toy problem. This suggests that in general, assigning an optimal $P^c$ may not be trivial for real systems. However, we reiterate that CMPC maintains a safe avoidance maneuver regardless of $P^c$ chosen; $P^c$ only tunes the performance of the system. In fact, CMPC with $P^c=0\%$ outperforms RMPC in the toy problem until the pop-up probability exceeds 84\%.

%The observant reader may note that $25\%$ is less than the expected pop-up probability of $1-(1-P)^X=$ \hbox{$1-(1-0.1)^{10}=65\%$}. However, the probability that an avoidance is required is smaller because late pop-ups do not breach $y=0$, reducing the avoidance probability to $1-(1-0.1)^{6}=47\%$. $25\%$ is lower still because mid-field pop-ups do not require a full deviation to $y=1$ (see Fig.~\ref{toy_cmpc_pop_soln}). These instances significantly reduce the $u^2$ penalty, and thus the optimal $P^c$ from what may be naively chosen based on triggering frequency. %$y^{obs}_{x=10} \leq 0.0$ if triggering occurs after $x=5$.

This toy example serves to illustrate the basic operation of the CMPC framework and demonstrates its performance advantage over a simple RMPC. Next, we demonstrate its effectiveness on an experimental platform.

 %%%%%%%%%%%%%%%%%%%%%%%%%%%%%%%%%%%%%%%%%%%%%%%

\section{Linear Contingency MPC for AVs} \label{section:cmpc4av} %%%%%%%%%%%%%%%%

%We now transition to demonstration of CMPC in a near-real world scenario. X1, a 2-ton student designed and built experimental research vehicle encounters a potential pop-up obstacle while driving down a narrow corridor. CMPC is employed to treat the potential obstacle as a contingent emergency

To implement CMPC on a full-size automated road vehicle, we adapt the linear formulation from Section~\ref{section:formulation} to the objectives and constraints specific to an AV system. The controller developed here is similar to that which we presented in \cite{Alsterda2019ContingencyVehicles}, in which Contingency MPC safely navigated an icy corner. The linearized dynamics model, its discretization, and optimization constraints were developed by Brown \textit{et al.}, and modified to solve directly for steering angle by Zhang \textit{et al.} \cite{Brown2017SafeVehicles}\cite{Zhang2018TireHandling}.

\subsection{Vehicle Dynamics Model}

\begin{figure}[t]
	\centering
    \includegraphics[trim={6.5cm 11.95cm 6.5cm 12cm },clip,scale=1]{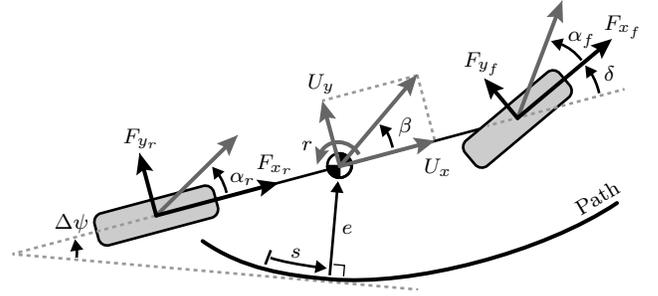} %LBRT
    \caption{Single track planar bicycle model diagram on a curvilinear path.}
    \label{bicycle}
\end{figure}

To handle contingency scenarios which push the limits of vehicle handling, dynamics must be modeled with an appropriate degree of fidelity. We accomplish this using a bicycle model with successively linearized tire forces \cite{Erlien2014IncorporatingControl}. Fig.~\ref{bicycle} illustrates the model with three position and velocity states, and tire forces at each axle. The position states are local to a path with curvature $\kappa$. $s$ represents the vehicle's longitudinal progress, $e$ its lateral error from the path, and $\Delta\psi$ is the heading angle error. Velocity states are composed of longitudinal speed $U_x$, lateral speed $U_y$, and yaw rate $r$. The following differential equations govern the states' evolution:
\vspace{-0.25 cm}

\begin{align*}
    \dot{s} &= U_x - U_y \Delta\psi \label{eq:s}\tag{6a}\\[0.3 cm]
    \dot{e} &=  U_y + U_x \Delta\psi \label{eq:e}\tag{6b}\\[0.25 cm]
    \dot{\Delta\psi} &= \, r \, - \, \kappa \, U_x \label{eq:dPsi}\tag{6c}\\[0.15 cm]
    \dot{U_x} &= \frac{F_{xf} + F_{xr}}{m} + r U_y \label{eq:Ux}\tag{6d}\\
    \dot{U_y} &= \frac{F_{yf} + F_{yr}}{m} - r U_x \label{eq:Uy} \tag{6e}\\
    \dot{r} &= \frac{aF_{yf} - bF_{yr}}{I_z} \label{eq:r} \tag{6f}
\end{align*}\vspace{-0.25cm}

%\begin{equation} \label{eq:s} \tag{6a}
%\dot{s} = U_x - U_y \Delta\psi \hspace{.6 cm}
%\end{equation}
%\vspace{.01 cm}
%\begin{equation} \label{eq:e} \tag{6b}
%\dot{e} =  U_y + U_x \Delta\psi \hspace{.6 cm}
%\end{equation}
%\vspace{.01 cm}
%\begin{equation} \label{eq:dPsi} \tag{6c}
%\dot{\Delta\psi} = \, r \, - \, \kappa \, U_x \hspace{1.3 cm}
%\end{equation}
%\vspace{0.01 cm}
%\begin{equation} \label{eq:Ux} \tag{6d}
%\dot{U_x} = \frac{F_{xf} + F_{xr}}{m} + r U_y
%\end{equation}
%\vspace{0.01 cm}
%\begin{equation} \label{eq:Uy} \tag{6e}
%\dot{U_y} = \frac{F_{yf} + F_{yr}}{m} - r U_x
%\end{equation}
%\vspace{0.01 cm}
%\begin{equation} \label{eq:r} \tag{6f}
%\dot{r} = \frac{aF_{yf} - bF_{yr}}{I_z} \hspace{.43 cm}
%\end{equation}
%\vspace{0.01 cm}

To linearize these equations for each iteration, the longitudinal trajectory is first computed upstream of CMPC by a simple feedforward-feedback controller, as developed by Funke \textit{et al.} \cite{Funke2017CollisionScenarios}. Therefore $F_{xf}, F_{xr}$, $U_x$, and $s$ are known constants to CMPC. With $F_x$ commands set, the linear CMPC formulated in this article may use only the steering angle $\delta$ to solve each optimization. Including longitudinal forces into CMPC is an opportunity for future development. The MPC state vector is then $x = [U_y \enspace r \enspace \Delta\psi \enspace e]^T$, and the command input $u = \delta$.

$F_{yf}$ and $F_{yr}$ are modeled by a nonlinear Fiala brush tire model that relates lateral tire forces to lumped slip angles $\alpha_f$ and $\alpha_r$, which are geometrically computed as follows \cite{Pacejka2012TireDynamics}:
\vspace{-0.1cm}

\begin{equation} \label{eq:af} \tag{7a}
\delta + \alpha_f = \tan^{-1}\left(\frac{U_y + ar}{U_x}\right) \hspace{.59 cm}
\end{equation}
\vspace{-.5 cm}

\begin{equation} \label{eq:ar} \tag{7b}
\alpha_r = \tan^{-1}\left(\frac{U_y - br}{U_x}\right)
\end{equation}
%\vspace{0.1cm}

\begin{figure}[t]
	\centering%\vspace{-5pt}
    \setlength{\fboxrule}{0pt}
    \framebox{\parbox{3.3in}{\centering \includegraphics[trim={0cm 0cm 0cm 0.1cm},clip,scale=0.9]{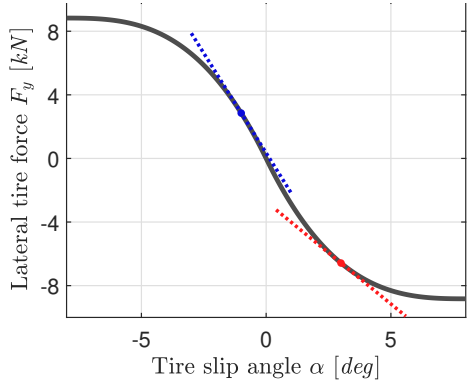}}} %LBRT
    \caption{Fiala brush tire model. Linearizations shown for two possible operating points, illustrating how nominal and contingency models may differ.}
    \label{fiala}
\end{figure}

\noindent The Fiala model is illustrated in Fig.~\ref{fiala}, accompanied by linearizations about two operating points. The curve is successively re-linearized before each CMPC iteration using the previous solution's states and inputs as operating points. Each stage among each horizon uses a unique linearization.

The linearized dynamics are then discretized with respect to time into twenty steps: five short 20 ms time-steps followed by fifteen longer 250 ms time-steps. The commanded steering angle $\delta$ is held constant with a zero-order hold during the short time-steps, and interpolated with a first order hold for the remainder of the horizon \cite{Brown2017SafeVehicles}. We chose the number and length of these time-steps to capture high frequency vehicle dynamics in the near-term, and to extend the total horizon far enough to plan appropriately. The CMPC control loop executes in less than one 20 ms time-step to ensure the next command is ready before the first expires. The resulting dynamics fit the affine form from equation~(\ref{eq:dyn_lin}), with block diagonal matrices to support both nominal and contingency prediction horizons.

%discrete affine dynamics take the form:
%\begin{equation} \label{eq:lin_dyn_old} \tag{4a}
%x_{k+1} = A_k x_k + B_k u_k + C_k
%\end{equation}
%To support both nominal and contingency prediction horizons, (\ref{eq:lin_dyn_old}) is duplicated for both $x^n$ and $x^c$ state vectors into a block diagonal format:

%\begin{align*} \label{eq:lin_dyn_c} 
%\textbf{x}_{k+1}
%= &\begin{bmatrix*}[c]
%    x^n \\ x^c
%\end{bmatrix*}_{k+1}  \tag{4b} \\[0.1 cm]
%= &\begin{bmatrix*}[c]
%    A^n & 0 \; \\ 0 & A^c \;
%\end{bmatrix*}_k \!\! \textbf{x}_k +
%\begin{bmatrix*}[c]
%    B^n & 0 \; \\ 0 & B^c \;
%\end{bmatrix*}_k \!\! \textbf{u}_k +
%\begin{bmatrix*}[c]
%    C^n \\ C^c
%\end{bmatrix*}_k 
%\end{align*}

\subsection{Linear CMPC Problem Statement} %%%%%%%%%%%%%%%%%%%%%%%%%%%%%%%%%%%%

The following CMPC optimization is extended from the deterministic MPC presented in \cite{Brown2017SafeVehicles} to calculate a smooth trajectory which follows a desired path while adhering to dynamics and environmental constraints:

\begin{align*} \label{eq:lin_J}
\underset{\textbf{u}}{\text{min}} \;\;
\sum_{k=0}^{20} \;\;
&\textbf{x}_k^\top
\begin{bmatrix} P^nQ & 0 \\ 0 & P^cQ \end{bmatrix}
\textbf{x}_k \tag{8a} \\[.1 cm]
+ \; &\textbf{v}_k^\top \,
\begin{bmatrix} P^nR & 0 \\[.1 cm] 0 & P^cR \end{bmatrix}
\textbf{v}_k \; + \;
W \sigma_k &
\end{align*} \vspace{-0.2 cm}

\begin{equation} \label{eq:lin_weights} \tag{8b}
\begin{aligned}
Q =
    \begin{bmatrix*}[c]
    \, 0 & 0 & 0 & 0 \, \\
  	\, 0 & 0 & 0 & 0 \, \\
    \, 0 & 0 & 1 & 0 \, \\
    \, 0 & 0 & 0 & 1 \, \\
 	\end{bmatrix*}
\quad &; \quad
R = 0.01 \\[10 pt]
\textbf{v}_k = 
	\begin{bmatrix*}[c]
    \, u_k^n - u_{k-1}^n \\
  	\, u_k^c - u_{k-1}^c \\
 	\end{bmatrix*} \hspace{.5 cm}
&; \quad
W = 1000
\end{aligned}
\end{equation} \vspace{-0.0 cm}

\noindent State weighting matrix $Q$ penalizes heading error $\Delta\psi$ and lateral error $e$. Input weight $R$ penalizes the slew rate of the steering angle $\delta_k - \delta_{k-1}$. $W$ heavily discourages violation of the environmental constraints in (\ref{eq:lin_envEnv}) by penalizing growth in the slack variable $\sigma$. Weight values were tuned for experimental performance, and to prioritize obstacle avoidance, path-tracking, and then smooth operation. This minimization is subject to the linearized dynamics and to the following inequality constraints. First, steering angle and slew rate limits are enforced:

\begin{equation} \label{eq:lin_uEnv} \tag{8c}
|\textbf{u}_k| \leq \begin{bmatrix*}[c] \delta_{max} \\ \delta_{max} \end{bmatrix*}
\quad \, ; \, \quad
|\textbf{v}_k| \leq \begin{bmatrix*}[c] v_{max} \\ v_{max} \end{bmatrix*}
\quad \forall \quad k
\end{equation}

\noindent Next, environmental and obstacle boundaries are enforced with slack, such that $e_{min} \leq e \leq e_{max}$: \vspace{-.2 cm}

\begin{align*} \label{eq:lin_envEnv}
&e^n_{min,k} - \sigma_k \, \leq \; e^n_k \; \leq \, e^n_{max,k} + \sigma_k \tag{8d} \\
&e^c_{min,k} - \sigma_k \, \leq \; e^c_k \; \leq \, e^c_{max,k} + \sigma_k \quad \forall \quad k
\end{align*}

\noindent Finally, trajectories are coupled via their first commands:

\begin{equation} \label{eq:lin_u0} \tag{8e}
u^n_0 = u^c_0
\end{equation}

 %%%%%%%%%%%%%%%%%%%%%%%%%%%%%%%%%%%%%%%%%%%%%%%%

\section{Experimental Results} \label{section:results} %%%%%%%%%%%%%%%%%%%%%%%%%%%%%%%%%%%%

\begin{figure*}
    \centering
	\includegraphics[width=\linewidth,trim={0.7cm 11.65cm 0.8cm 11.75cm},clip]{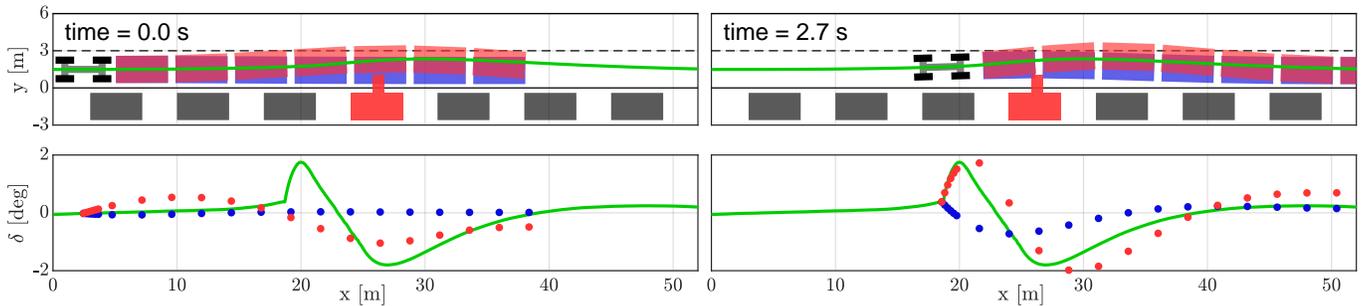}%LBRT
    \caption{CMPC safely navigates around a car door opening into its lane. The contingency horizon (transparent red) maintains the feasibility of an avoidance maneuver, while the nominal horizon (transparent blue) focuses on the desired path. Closed loop position $(x,y)$ and steering angle $\delta$ are plotted in green. The parked car's door begins to open just after 2.7 sec and the avoidance maneuver is deployed.}
    \label{exp_ph}
\end{figure*}

%\begin{figure*}\centering
%	\includegraphics[width=\linewidth]{cardoor_PH_plots.eps}\end{figure*}

\begin{figure}[!b]
    \centering
	\includegraphics[width=\linewidth - 0.0 cm]{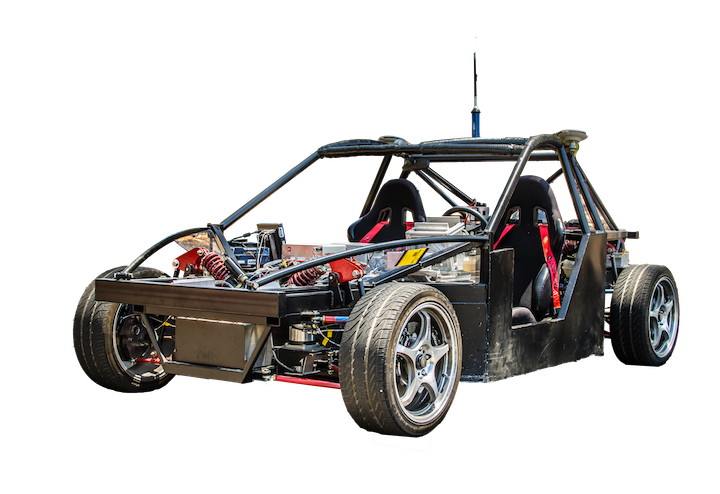}
	% [trim={0cm 0cm 0cm 0cm},clip,scale=.7] LBRT
    \caption{X1 is a modular experimental research platform with automated driving capabilities at Stanford University.}
    \label{x1}
\end{figure}

Using the formulation defined in Section~\ref{section:cmpc4av}, we validated CMPC experimentally in a realistic emergency scenario for an automated road vehicle. CMPC safely controlled X1, the 2-ton research vehicle shown in Fig.~\ref{x1}, down a narrow lane of the two-way street illustrated in Fig.~\ref{exp_ph}, with parallel-parked cars lining X1's right-hand side. One parked car, illustrated in red near $x=24$ m, pulled into its parking spot very recently. Therefore its occupants are likely to exit the vehicle, and may open their car door into X1's intended path. The potential for X1 to collide with the car door is thus a contingent emergency.

X1 is an electrically powered AV research platform equipped with steer-, \mbox{drive-}, and brake-by-wire systems to enable automated  experiments. X1 measures its pose, velocity, and acceleration with a Novatel dGPS/INS with RTK. Low-level control is computed on a real-time dSpace MicroAutoBox executing at 500 Hz. The CMPC optimization was computed on an i7 x86 CPU using CVXGEN in a Linux C++ environment at 50Hz \cite{Mattingley2012CVXGEN:Optimization}. No visual perception sensors were used in this experiment; the road geometry and obstacle description were defined in GPS coordinates and made known to CMPC \textit{a priori}. To incorporate CMPC into a hypothetical commercial system, it will be necessary to perceive and identify contingent hazards, a task beyond the scope of this article.

Fig.~\ref{exp_ph} illustrates two snapshots in time as X1 approaches the recently parked red vehicle at a speed of 12 m/s, about 27 mph. Both snapshots are taken before the car door opens; its \textit{potential} footprint is illustrated in red, extending into X1's lane. The car door is forecast to have a maximum width of \mbox{1 m} and an opening speed of 2 m/s, both known to CMPC and encoded into the contingency horizon's $e^c_{min}$. In this test The CMPC likelihood parameter was set to $P^c=25\%$, indicating the contingency has a moderate likelihood.

In most real-life occurrences of this scenario, the parked car's occupants will act reasonably, exiting their vehicle after X1 has passed. However, it's possible the door could open just as X1 arrives and create an emergency which requires anticipatory behavior to avoid. A na\"ive controller planning to react if and when the door opens may not be able to escape collision. CMPC acts robustly to ensure the feasibility of escape, but balances conservatism with an incentive to follow the nominal path.

At top-left, the position and heading states of the CMPC prediction horizons are plotted at $t=0$ sec. The contingency horizon suggests a smooth avoidance around the door's potential footprint. The nominal states do not see the potential hazard and are free to pass through. To avoid clutter, only seven states are drawn from each horizon. The corresponding steering commands are plotted in the lower-left. These command trajectories share an identical steering angle at $u_0$, where for the time being they agree to adhere to the lane's center-line.

At top-right X1 has closed the distance to the potential obstacle considerably, but the car door remains closed for now. X1's closed-loop positions and steering angles (green) leading up to this point illustrate an important behavior: To reduce the likelihood of requiring drastic emergency behavior, CMPC has guided X1 off the lane's center-line and away from the potential hazard by $\small \sim \! 11$ cm. The magnitude of this conservative deviation is controlled by $P^c$.

The difference between the blue and red horizons at this point illustrates significant tension built up in CMPC's cost function. The nominal horizon wants to return to the path, but the contingency horizon ensures a moderately smooth avoidance is available. X1's trajectory leading up to this snapshot shows the history of negotiation between these objectives.

%\begin{figure}[!b]%[hb]\centering
%	\includegraphics[width=\linewidth-0.5cm]{cardoor_CL_plots.eps}\end{figure}

\begin{figure}[!b]
    \centering
	\includegraphics[width=\linewidth-0.5cm,trim={6.0cm 9.0cm 6.0cm 9.0cm},clip]{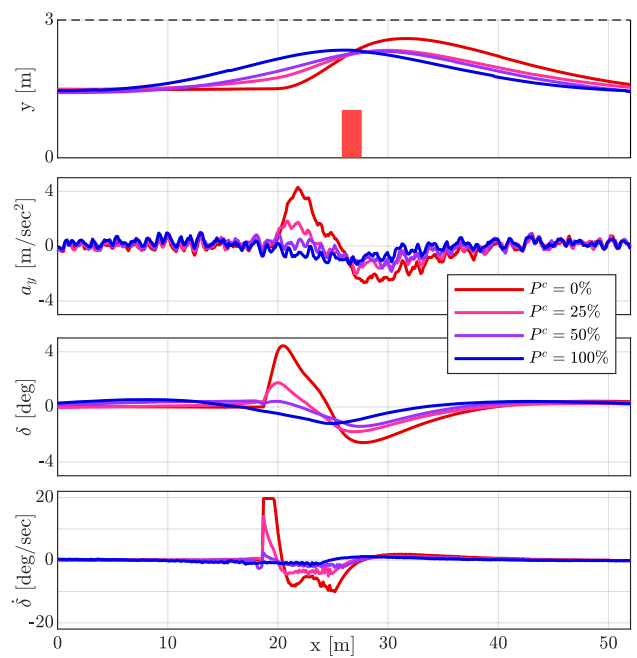}%LBRT
    \caption{Closed-loop trajectories for CMPC navigating the car door scenarios from Fig.~\ref{exp_ph} with various $P^c$ settings. Plotted from the top: Zoomed position, lateral acceleration, steering angle, and steering rate.}
    \label{exp_cl}
\end{figure}

In the very \textit{next} time-step after 2.7 sec, the car door begins to swing open. At this moment the contingency is no longer a possibility, but reality. CMPC's understanding the of the situation should evolve when a contingency occurs, and there are several design choices available to encode this change in circumstance. In this experiment, CMPC is made aware of the opening door by including the door's footprint into the nominal horizon's constraints (previously, it was only visible to the contingency horizon). With both horizons now aware of the emergency, the tension between plans is resolved and X1 steers purposefully around the obstacle.

Next, Fig.~\ref{exp_cl} shows the closed loop behavior of CMPC navigating the car door emergency with a range of $P^c$ values. In each experiment, the parked car's door triggered to open at the same time. The $P^c$ sweep illustrates the how this parameter affects CMPC's approach and reaction to emergency. At $0\%$, CMPC does not deviate from the desired path until the car door starts to open. This reaction requires the most lateral acceleration and steering angle, and saturates the model's slew rate limit. As $P^c$ is increases, CMPC takes an increasingly conservative approach, eventually performing a worst-case Robust MPC avoidance when $P^c=100\%$.

 %%%%%%%%%%%%%%%%%%%%%%%%%%%%%%%%%%%%%%

\section{Discussion} \label{section:discussion} %%%%%%%%%%%%%%%%%%%%%%%%%%%%%%%%%%%%%%%%%%%%%%%%%%%%%%%%%%%

The toy simulation and car-door experiments in this article demonstrate that Contingency MPC can navigate uncertain environments with known potential hazards. The type of hazard presented here was a pop-out obstacle, but CMPC can also manage model mismatch in the form of friction uncertainty \cite{Alsterda2019ContingencyVehicles}.

Mathematically, the hazards we investigated represent two broad classes of contingencies that can be encoded into CMPC. First, the inequality constraints can inscribe obstacles like the car door. Other hazards encoded this way include command limits such as a vehicle's maximum steering angle or braking force. Second, the equality constraints can be modified to encode model mismatch. Uncertainty over any parameter in the system dynamics can constitute a contingency.

% One class not yet investigated, however, are contingencies in the optimization's cost function; this remains an opportunity for further investigation.

CMPC offers a flexible structure for robust control, with several design choices left to the control engineer. Selecting a value for $P^c$ has important implications for the system's performance, as summarized in Table \ref{tab:Pc}. Upper-left and lower-right boxes describe the intended CMPC design.
\vspace{0.2 cm}

\begin{table}[ht]
\centering
\caption{\label{tab:Pc}}
\begin{tabular}{Sc|Sc|Sc|}
    &
    \makecell{Contingent event\\is common} &
    \makecell{Contingent event\\is uncommon} \\
    \hline
    Large $P^c$ &
    \makecell{\greencheck More anticipatory \\ deviation; less \\ dynamic avoidances} &
    \makecell{$\redxmark$ More anticipatory \\ deviation, often \\ unnecessary}\\
    \hline
    Small $P^c$ &
    \makecell{$\redxmark$ Less anticipatory \\ deviation; more frequent \\ dynamic avoidances} &
    \makecell{\greencheck Less anticipatory \\ deviation; less frequent \\ dynamic avoidances} \\
    \hline
\end{tabular}
\end{table}
\vspace{0.2 cm}

If a contingent event is common, its horizon should have greater cost burden, imparted by a larger $P^c$. This ensures smooth avoidances are maintained for these frequent events. Conversely, uncommon contingencies should have a smaller cost burden, directing CMPC to pursue higher nominal performance. When the contingency does occur, a more dynamic avoidance may be used. It is critical to note that CMPC remains robust to the contingency regardless of the value assigned to $P^c$.

The effect of this design is clear in the context of the car door experiment. If car door pop-outs rarely occur, an avoidance will seldom be deployed. Therefore CMPC should focus on nominal operation and allow the contingency plan freedom to use higher slew rate. Alternatively, if pop-outs are more common, deployment of the contingency plan will be more frequent. In that case the planned avoidances should use less slew, but maintaining them will require more compromise to nominal objectives (e.g. path following or speed tracking).

Another design choice considers how to react when contingencies occur. In this article, pop-out detection was communicated to the nominal horizon; hence both plans perform the avoidance. Alternatively, $P^c$ could be updated online, or CMPC could be collapsed to a single-horizon MPC when emergency occurs. Choice among these options depends on the control application and emergency circumstance.

An opportunity for development is to include throttle and brakes into CMPC. In this article the convex optimization could only use steering to mitigate the hazard. In practice, however, the best solution to reduce risk may be to simply slow down. Longitudinal commands could be incorporated by encoding the full nonlinear vehicle dynamics into CMPC.

One caveat for obstacle avoidance should be stated: Obstacles that can appear or pop out instantaneously may not be appropriate for CMPC. In the case of the car door, CMPC leverages knowledge of the obstacle's maximum pop-out speed to calculate how closely it can pass by an unopened door. If the door can open arbitrarily fast, there is no safe buffer distance except to clear the obstacle's entire potential footprint. CMPC is not necessarily suitable for every potential hazard.

Finally, Bloom and Menefee remind us that not all contingencies are negative \cite{Bloom1994ScenarioPlanning}. A positive contingent event may offer a slim passing opportunity to a race car, but only if the controller is poised with a plan to take advantage.

\section{Conclusion} \label{section:conclusion} %%%%%%%%%%%%%%%%%%%%%%%%%%%%%%%%%%%%%%%%%%%%%%%%%%%%%%%%%%%

In this article, Contingency Model Predictive Control is established as a credible strategy to augment a deterministic linear MPC controller with robustness. In systems where potential emergencies can be identified, CMPC maintains an avoidance trajectory while pursing performance objectives to the greatest extent possible. Experimentally, the controller successfully navigated a real-world obstacle avoidance scenario with an intuitive approach that achieved higher performance than a worst-case Robust MPC.

Promising avenues for future development of this research include: 1) applying the CMPC framework to new applications outside vehicle automation, and to other AV scenarios such as pedestrian avoidance or safely following vehicles that are liable to stop (e.g. mail or garbage trucks); 2) integrating CMPC into a real-time emergency recognition system or a visual perception system, capable of identifying contingencies; and 3) considering multiple contingent events simultaneously, which may require multiple contingency horizons.

\newpage

 %%%%%%%%%%%%%%%%%%%%%%%%%%%%%%%%%%%%%

\appendix[Explicit Solutions to Toy Problems] \label{section:appendix}

% or

% use appendices with more than one appendix then use \section to start each
% appendix you must declare a \section before using any \subsection or using
% \label (\appendices by itself starts a section numbered zero.)

\subsection{Robust MPC Solution} %%%%%%%%%%%%%%%%%%%%%%%

An analytic solution for the Toy Problem RMPC optimization follows. The potential obstacle is positioned at the end of the horizon at $k=N$. The problem statement is: \vspace{0 cm}

\begin{equation*} \label{eq:uni_r_j} \tag{9a}
    \min\limits_{\textbf{u}} \; \sum_{k=0}^{N} \; u_k^2
    = \min\limits_{\textbf{u}} \; \textbf{u}^\top \textbf{u}
\end{equation*} \vspace{0 cm}

\noindent subject to dynamics: \vspace{-.2 cm}

\begin{equation*} \label{eq:uni_r_dyn} \tag{9b} \begin{aligned}
    y_N \, = \, y_{N-1} + u_{N-1}
    \, = \, \cancelto{0}{y_0} \, + \, \sum_{k=0}^{N-1} u_k
    \, = \, \mathbbm{1}^\top \textbf{u}
\end{aligned} \end{equation*} \vspace{0 cm}

\noindent and obstacle constraint: \vspace{0 cm}

\begin{equation} \label{eq:uni_r_obs} \tag{9c}
    y_N = \mathbbm{1}^\top \textbf{u} \geq y^{obs}
\end{equation} \vspace{-.1 cm}

\noindent This optimization is solved by Lagrange multipliers:

\begin{equation*} \begin{aligned}
    &\Lagr(\textbf{u},\lambda) = \textbf{u}^\top \textbf{u} - \lambda (\mathbbm{1}^\top \textbf{u} - y^{obs}) \\[.5 cm]
    \triangledown_{\lambda}&\Lagr(\textbf{u},\lambda)
    = 0 = \mathbbm{1}^\top \textbf{u} - y^{obs} \quad \; \rightarrow \quad \mathbbm{1}^\top \textbf{u} = y^{obs} \\[.2 cm]
    \triangledown_{\textbf{u}}&\Lagr(\textbf{u},\lambda)
    = \vec{0} = 2 \, \textbf{u} - \lambda \cdot \mathbbm{1} \hspace{.6 cm} \rightarrow \quad u_* = \lambda / 2 \\[.5 cm]
    &\rightarrow u_0 = u_{N-1} = y^{obs}/N
\end{aligned} \end{equation*} \vspace{-.4 cm}

\begin{flalign} \label{eq:rmpc_sol} \tag{9d}
    \text{RMPC Solution: }
    \hspace{30 pt}
    \boxed{u_0 = y^{obs}/N} &&
\end{flalign} \vspace{0 cm}

This analytic solution is employed by the toy problem simulations in Section~\ref{section:toy}. When the potential obstacle enters into the purview of RMPC, it takes immediate action to avoid it, amortizing its inputs $u_*$ evenly across the horizon.

\subsection{Contingency MPC Solution} %%%%%%%%%%%%%%%%%%

We follow a similar strategy to solve CMPC. The optimization problem statement is:

\begin{equation*} \label{eq:uni_c_j} \tag{10a} \begin{aligned}
    &\min_{\textbf{u}} \; \sum_{k=0}^{N} \hspace{8 pt}
    \begin{bmatrix*}[c] \, u^n \, \\ \, u^c \, \end{bmatrix*}_k^T
    \begin{bmatrix*}[c] \, P^n & 0 \, \\ \, 0 & P^c \, \end{bmatrix*}
    \begin{bmatrix*}[c] \, u^n \, \\ \, u^c \, \end{bmatrix*}_k \\[.3 cm]
    = &\min_{\textbf{u}} \quad \textbf{U}^T \textbf{P} \, \textbf{U}
\end{aligned} \end{equation*} \vspace{0 cm}

\noindent where $\; \textbf{U} = [ \; u_0 \; , \; u^n_1 \cdots u^n_N \; , \; u^c_1 \cdots u^c_N ] \; \in \; \R^{2 N -1}$, concatenating the entire horizon of nominal and contingency inputs. $\textbf{P}$ is the following matrix $\in \R^{(2 N -1)\text{x}(2 N -1)}$:

\begin{equation*}
    \begin{bmatrix*}
        1.0 & 0                    & 0                    \\[0 pt]
        0   & P^n \cdot \textbf{I} & \textbf{0}           \\[0 pt]
        0   & \textbf{0}           & P^c \cdot \textbf{I}
    \end{bmatrix*}
\end{equation*} \vspace{-.1 cm}

\noindent \textbf{I} and \textbf{0} are the identity and zero matrix, respectively, $\in \R^{(N-1)\cdot(N-1)}$. $P^n + P^c = 1.0$ in the top-left position representing the total weight on $u_0$, which replaced $u^n_0$ and $u^c_0$ in (\ref{eq:uni_c_j}). The optimization is subject to the dynamics equations: \vspace{-.1 cm}

\begin{equation*} \label{eq:uni_c_dyn} \tag{10b} \begin{aligned}
    y^n_N &= u_0 + \sum_{k=1}^{N-1} u^n_k 
    = {\mathbbm{1}^n}^\top \textbf{U} \\[.2 cm]
    y^c_N &= u_0 + \sum_{k=1}^{N-1} u^c_k
    = {\mathbbm{1}^c}^\top \textbf{U}
\end{aligned} \end{equation*} \vspace{0 cm}

\noindent ${\mathbbm{1}^n}^\top \! = [ \, 1 \; 1 \! \mydots \! 1 \; 0 \! \mydots \! 0 \, ]$ and ${\mathbbm{1}^c}^\top \! = [ \, 1 \; 0 \! \mydots \! 0 \; 1 \! \mydots \! 1 \, ] \, \in \, \R^{2 N \scalebox{1.2}[1]{-} 1}$. The constraints are: \vspace{-0.3 cm}
%similar for $\mathbbm{1}^n$.

\begin{equation} \label{eq:uni_c_u0} \tag{10c}
    u_0 = u^n_0 = u^c_0
\end{equation} \vspace{-0.7 cm}

\begin{equation} \label{eq:uni_c_obs} \tag{10d}
    y^c_N = {\mathbbm{1}^c}^\top \textbf{U} \geq y^{obs}
\end{equation} \vspace{-0.3 cm}

\noindent Once again, solution by Lagrange multiplication: %(\ref{eq:uni_c_u0}) was baked into the reformulation of (\ref{eq:uni_c_j}) using $\textbf{U}$ and $\textbf{P}$.
\vspace{-.2 cm}

\begin{alignat*}{3}
    &\Lagr(\textbf{U},\lambda) = \textbf{U}^\top \textbf{P} \, \textbf{U} - \lambda ({\mathbbm{1}^c}^\top \textbf{U} &&- y^{obs}&&) \\[.5 cm]
    \triangledown_{\lambda}&\Lagr(\textbf{U},\lambda) = 0 = {\mathbbm{1}^c}^\top \textbf{U} - y^{obs}
    \hspace{0 cm} &&\rightarrow \hspace{0 cm} &&{\mathbbm{1}^c}^\top \textbf{U} = y^{obs} \\[.2 cm]
    \triangledown_{\textbf{U}}&\Lagr(\textbf{U},\lambda) = \vec{0} = 2 \, \textbf{U} - \lambda \cdot \mathbbm{1}^c
    \hspace{0 cm} &&\rightarrow \hspace{0 cm} &&u_0 = \lambda / 2\\[.15 cm]
    &&&\text{and} &&u^c_{k>0} = \frac{\lambda}{2 P^c} \\[.2 cm]
    &&&\text{and} &&u^n_{k>0} = 0
\end{alignat*}

Substituting for $\lambda$, we find the following CMPC solution for $u_0$ as a function of $P^c$:
\vspace{-0.1 cm}

\begin{empheq}[box=\fbox]{align}
    u_0 \, &= \, u^c_{k>0} \cdot P^c \nonumber \, = \, y^{obs} \frac{P^c}{P^c + N -1} \nonumber \\[5 pt]
    &= y^{obs}/N \hspace{12 pt} \text{if} \quad P^c = 100\% \tag{10e} \\[5 pt]
    &= 0 \hspace{37 pt} \text{if} \quad P^c = 0\% \nonumber
\end{empheq} %\vspace{0 cm}

\begin{figure}[!t]
    \centering
	\includegraphics[width=5.8 cm]{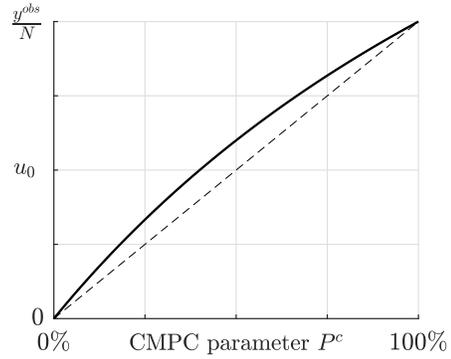}
	% [trim={0cm 0cm 0cm 0cm},clip,scale=.7] LBRT
    \caption{$u_0$ returned by CMPC, as a function of $P^c$, for a toy problem. The concave-down shape is shown in contrast to a linear gain (dashed line).}
    \label{uni_u0_Pc}
\end{figure}

As illustrated in Fig.~\ref{toy_cmpc_soln}, CMPC takes the minimum action necessary to maintain safety when $P^c=0\%$. As $P^c$ increases, CMPC takes an increasingly conservative approach. $u_0$ goes up monotonically with the concave-down shape shown in Fig.~\ref{uni_u0_Pc} until it fully adopts RMPC behavior at $P^c=100\%$.

%Thus far, the relationship between Contingency and Robust MPC has been shown only for the narrow toy problem, but the equivalency of the two programs when $P^c=100\%$ extends to any system. A generic RMPC problem statement can written:

%\begin{flalign*}
%    \min\limits_{u} \, &\sum_{k=0}^{N} \; j(x_k,u_k) &&\\
%    \text{such that:} \hspace{1.8 cm}&&& \\
%    x_{k+1} &= f(x_k,u_k) \hspace{.4cm} \forall \hspace{.4cm} k &&\\
%    g(x,u) &\leq 0 \hspace{1.6 cm} \forall \hspace{.4cm} g &&
%\end{flalign*} \vspace{.2 cm}

%\noindent By inspection, the CMPC formulation from equations (1a-d) reduce to same problem statement when $P^c=100\%$, assuming contingency constraints are more tight than the nominal case. The contingency states and inputs are now the only meaningful variables, tantamount to solving a Robust MPC.

%\begin{flalign*}
%    \min\limits_{u^n,u^c} \, &\sum_{k=0}^{N} \; \cancelto{0}{P^n} \cdot j(x_k^n,u_k^n) + \cancelto{1}{P^c} \cdot j(x_k^c,u_k^c) &&\\
%    = \min\limits_{u^c} \, &\sum_{k=0}^{N} \; j(x_k^c,u_k^c) &&\\
%    \text{such that:} \hspace{1.8 cm} &&&\\
%    x^c_{k+1} &= f^c(x^c_k,u^c_k) \hspace{.4cm} \forall \hspace{.4cm} k &&\\
%    g(x^c,u^c) &\leq 0 \hspace{1.75cm} \forall \hspace{.4cm} g &&
%\end{flalign*} \vspace{.2 cm}

%\begin{equation*}
%    R \in \R^+ \; \text \quad \text{and} \quad \textbf{R} = diag(R) \in \R^{N\text{x}N}
%\end{equation*} \vspace{.1 cm}

 %%%%%%%%%%%%%%%%%%%%%%%%%%%%%%%%%%%%%%%%%%%%%%%%%%

\section*{Acknowledgment} % use section* %%%%%%%%%%%%%%%%%%%%%%%%%%%%%%%%%%%%%%%

The authors would like to thank Renault Group, VW Group Research, and VW ERL for experimental support. Alsterda is supported by Ford Motor Company and the U.S. Dept. of Veterans Affairs G.I. Bill.

% References Section %%%%%%%%%%%%%%%%%%%%%%%%%%%%%%%%%%%%%%%%%%%%%%%%%%%%%%%%%%%

\bibliographystyle{IEEEtran}
\bibliography{Sections/8_references}

%\bibliographystyle{IEEEtran}
%bibliography{IEEEabrv,../bib/paper}
%\begin{thebibliography}{1}
%\end{thebibliography}

% To put refs on page by themselves when using endfloat & captionsoff option:
\ifCLASSOPTIONcaptionsoff
  \newpage
\fi

% trigger a \newpage just before the given reference number - used to balance
% the columns on the last page adjust value as needed - may need to be
% readjusted if the document is modified later: \IEEEtriggeratref{8}
% "triggered" cmd if desired: \IEEEtriggercmd{\enlargethispage{-5in}}


\vspace{-.4 cm}

\begin{IEEEbiography}[{\includegraphics[width=1in, height=1.25in, trim=60 0 120 0, clip, keepaspectratio]{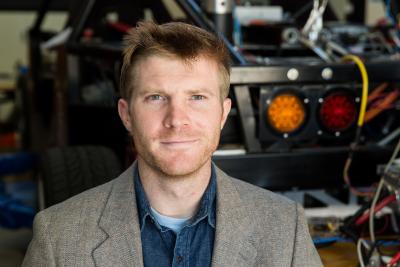}}]{John P. Alsterda} %LBRT
received the M.S. degree in mechanical engineering from Stanford University, Stanford, CA, USA, in 2018, and the B.S. degree in Physics from University of Illinois, Urbana Champaign, IL, USA, in 2011. He is currently pursuing th Ph.D. degree with Stanford University, Stanford, CA, USA. His current research interests include path planning and control for automated systems under uncertainty. He is a Lt.Cdr. in the United States Naval Reserve at the Office of Naval Research.
\end{IEEEbiography}

\vspace{-.4 cm}

\begin{IEEEbiography}[{\includegraphics[width=1in,height=1.25in,trim=0 0 10 0,clip,keepaspectratio]{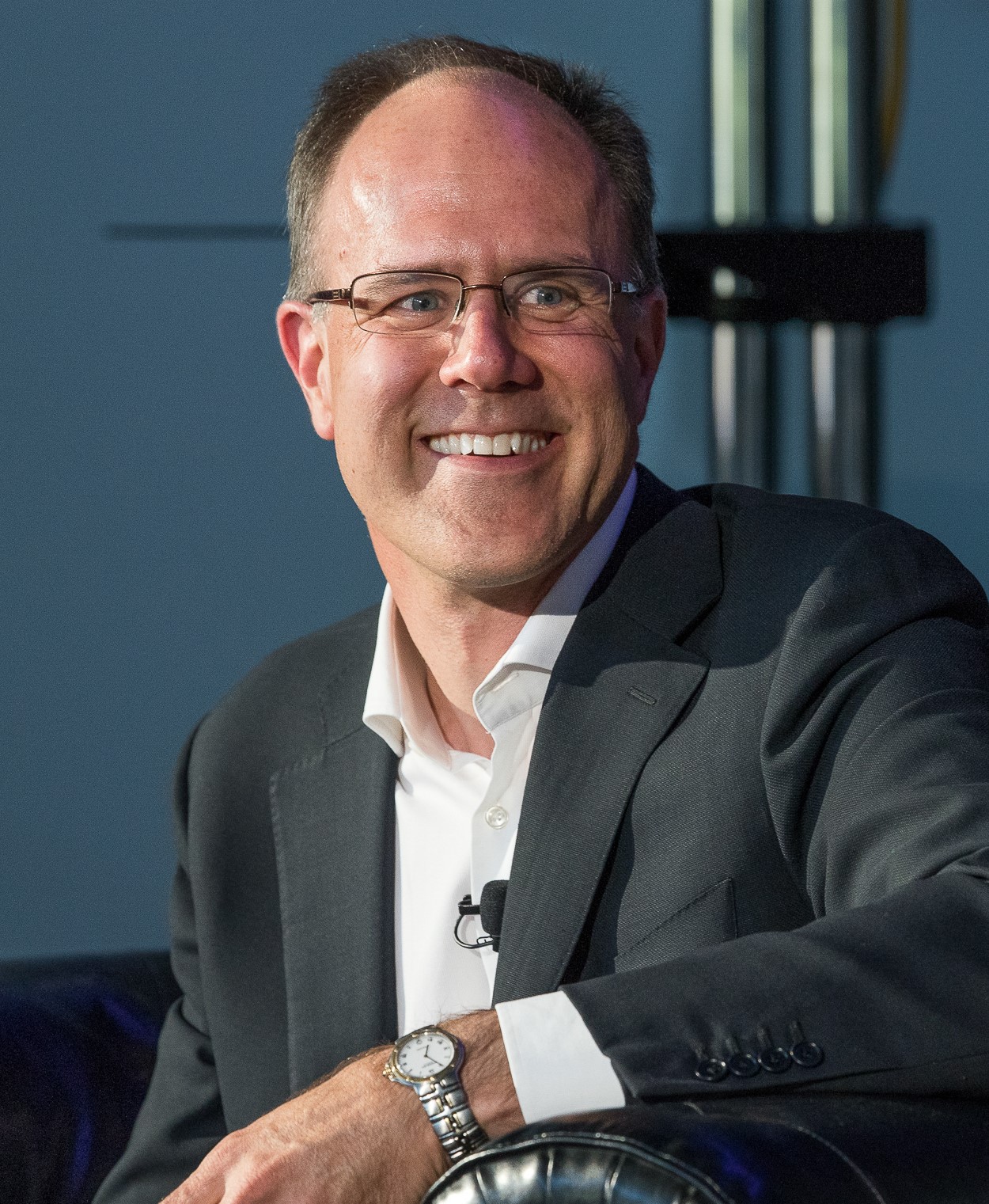}}]{J. Christian Gerdes} %LBRT
received the Ph.D. degree from the University of California at Berkeley, Berkeley,
CA, USA, in 1996. He is currently a Professor of mechanical engineering at Stanford University, Stanford,CA, USA, and the Director of the Center for Automotive Research at Stanford, Stanford University. He is a Co-Founder of Peloton Technology, Mountain View, CA, USA. His laboratory studies how cars move, how humans drive cars, and how to design future cars that work cooperatively with the driver or drive themselves. When not teaching on campus, he can often be found at the racetrack with students, instrumenting historic race cars, or trying out their latest prototypes
for the future.
Prof. Gerdes and his team have been recognized with several awards, including
the Presidential Early Career Award for Scientists and Engineers, the Ralph
Teetor Award from SAE International, and the Rudolf Kalman Award from the
American Society of Mechanical Engineers.
\end{IEEEbiography}

\end{document}